# Remarkable Enhancement of High Harmonic Generation from Superhard Material under High Pressure

Zishao Wang,[1,§] Tong Wu,[1,§] Ziwen Wang,[1,§] Sicheng Liu,[1] Hui Li,[1] Kun Zhao,[1] Jian Sun,[2,*] Chao Yu,[1,†] and Ruifeng Lu[1,‡]

[1]*Institute of Ultrafast Optical Physics, Department of Applied Physics & MIIT Key Laboratory of Semiconductor Microstructure and Quantum Sensing, Nanjing University of Science and Technology, Nanjing 210094, China*
[2]*National Laboratory of Solid State Microstructures, School of Physics and Collaborative Innovation Center of Advanced Microstructures, Nanjing University, Nanjing 210093, China*

High harmonic generation (HHG) in solids offers a pathway to develop compact extreme ultraviolet (EUV) sources crucial for attosecond science and advanced spectroscopy. Here, we demonstrate theoretically that high pressure dramatically enhances HHG in superhard hexagonal tungsten nitride ($h$-$WN_6$). Compared to solid state systems at ambient pressure, the reshaped electronic environment under high pressure leads to a unique bandgap widening of $h$-$WN_6$, which raises the material's damage threshold, allowing the utilization of stronger laser fields and enabling access to higher energy bands. This pressure-induced confinement is hopeful as a novel strategy to overcome the cutoff limitation for solid-state EUV light sources.

High harmonic generation (HHG) is a cornerstone of ultrafast nonlinear optics, enabling the production of coherent extreme ultraviolet (EUV) to soft X-ray radiation and attosecond pulses, which are indispensable for time-resolving electron dynamics in atoms, molecules, and condensed matter [1–4]. Initially demonstrated and extensively studied in gases, HHG in solids has emerged as a highly promising alternative [5,6], offering advantages such as higher atom density, significantly lower driving laser intensity requirements, and the potential for on-chip integration [7,8]. Also, solid-state HHG is a powerful tool for all-optical probing of material properties, including electronic band structure, crystal symmetry, and ultrafast carrier dynamics [9–11].

A critical performance metric for HHG is the maximum achievable photon energy, or cutoff energy [12]. Extending this cutoff to higher energies is paramount for various applications. For instance, higher photon energies facilitate the generation of shorter attosecond pulses, pushing the frontiers of temporal resolution in probing electron motion [13]. In spectroscopy, HHG serves as the engine for advanced techniques like attosecond transient absorption spectroscopy [14,15] and multidimensional spectroscopy [16,17] enabling real-time observation of ultrafast electron dynamics in complex molecules. Furthermore, in coherent diffractive imaging, shorter wavelengths directly translate to improved spatial resolution [18]. Despite considerable efforts to extend the cutoff such as driving-wavelength selection [19], crystal-orientation control [20], and heterostructure design [21], further extending the cutoff to higher energies generally requires stronger driving fields, and is therefore ultimately limited by the material's optical damage threshold. Overcoming this limitation in a robust and widely applicable manner remains a significant challenge.

External stimuli that can controllably modify material properties provide new avenues for optimizing HHG [22,23]. High pressure regime in solids can profoundly alter both crystal and electronic structures, leading to novel material phases and functionalities [24]. While pressure effects on linear optical properties are well-documented, its application to tailor strong-field nonlinear phenomena like HHG in solids is largely unexplored. Experimental studies on HHG in solids under high static pressure are particularly scarce.

A nitrogen-rich tungsten nitride ($h$-$WN_6$), first theoretically predicted and subsequently synthesized, is a superhard material exhibiting remarkable stability under extreme pressures exceeding 40 GPa [25]. Notably, $h$-$WN_6$ displays an anomalous increase in its bandgap with applied pressure—a characteristic attributed to enhanced repulsion of nitrogen lone-pair electrons within its unique ring structures under compression [26]. As demonstrated below, this property proves highly beneficial for HHG.

In this work, we calculated the critical electron density of $h$-$WN_6$ under various pressures. Previously, Stampfli and Bennemann showed that a sufficiently high density of conduction band electrons excited by an intense laser pulse can weaken chemical bonding and induce lattice instability, which is manifested by phonon softening and the appearance of imaginary phonon frequencies [27].

Following this physical picture, we use the phonon spectrum at different electronic excitation densities for evaluating the onset of laser-induced structural instability. In practice, we model this effect by reducing the total number of valence electrons in phonon spectrum calculation, which in fact considers electronic excitations to all unoccupied bands. Once the excited electron density exceeds a critical level, the initial lattice becomes unstable, accompanied by phonon softening. The phonon frequency ($\omega$) was determined by solving the characteristic equation for the dynamical matrix ($D_{ij}(q)$) at wave vector $q$:

$$\det\left|D_{ij}(\mathbf{q}) - \omega^2\delta_{ij}\right| = 0 \qquad (1)$$

Structural instability is indicated by the appearance of imaginary frequencies ($\omega^2<0$) in the phonon spectrum. The critical electron density, at which imaginary frequencies (soft modes) appear in the phonon spectrum (signifying dynamical instability and structural softening), is identified as the precursor to damage.

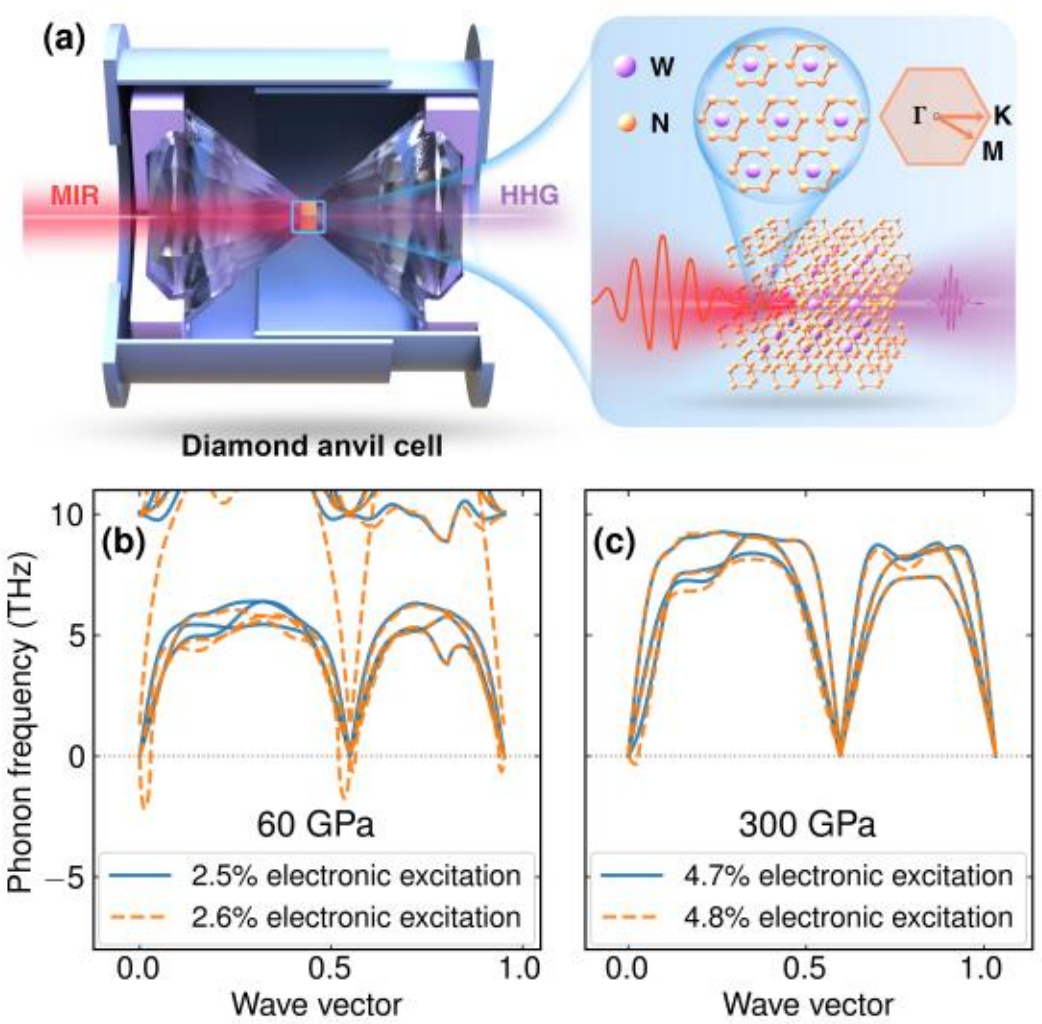

FIG. 1. HHG and stability analysis of high-pressure $h$-$WN_6$. (a) Schematic of the experimental apparatus for generating high harmonics from $h$-$WN_6$ pressurized in a diamond anvil cell. Calculated phonon dispersion at (b) 60 GPa and (c) 300 GPa under electronic excitation.

The ground-state electronic structure of $h$-$WN_6$ under various pressures (ranging from 60 GPa to 300 GPa) was calculated using density functional theory (DFT) as implemented in the Vienna Ab initio Simulation Package [28]. We employed the generalized gradient approximation with the Perdew-Burke-Ernzerhof functional for exchange and correlation. A plane-wave energy up to 1200 eV and a Monkhorst-Pack $k$-point mesh of 19×19×19 were utilized to ensure convergence of the total energy. The crystal structures at each pressure were fully relaxed until the energy and force are less than $10^{-5}$ eV and 0.01 eV/Å, respectively. To accurately capture the band structure for subsequent dynamical simulations, maximally localized Wannier functions were constructed using the Wannier90 package [29]. The W 5d orbitals and N 2s/2p orbitals were included in the projection to build the Wannier basis. Moreover, *ab initio* molecular dynamics simulations in an NPT ensemble were performed to assess the finite-temperature structural stability of $h$-$WN_6$ (see Figs. S1-S3 in Supplemental Material).

Fig. 1(a) shows a schematic of a potential experimental setup employing a diamond anvil cell, which facilitates high pressure applied to $h$-$WN_6$ while an intense incident laser drives the HHG process. The high-pressure phase of $h$-$WN_6$ is anticipated to allow much more electronic excitation in strong laser fields. To quantify this, we determine the critical (the highest) density of electrons excited from the valence band to the conduction band at which the crystal lattice becomes dynamically unstable. This stability analysis is based on phonon dispersion with a certain electronic excitation percentage. For instance, at 60 GPa, the $h$-$WN_6$ structure is stable at electronic excitation percentage of 2.5% but exhibits clear imaginary frequencies at 2.6% electronic excitation, indicating the onset of instability, as shown in Fig. 1(b). At higher pressure of 300 GPa, 4.8% electronic excitation results in structural instability, as seen from Fig. 1(c).

The evolution of electronic density was simulated by numerically solving the semiconductor Bloch equations (SBEs) [30–32], taking into account both interband polarization and intraband oscillation. The time evolution of the density matrix elements $\rho_{nm}(k)$ (representing coherence between bands $n$ and $m$, or population if $n = m$) at crystal momentum $k$ under the influence of the laser electric field $E(t)$ is governed by:

$$i\,\partial_t \rho_{nm}^{k(t)} = \left[\varepsilon_m^{k(t)} - \varepsilon_n^{k(t)} - \frac{\mathrm{i}(1-\delta_{mn})}{T_2}\right]\rho_{nm}^{k(t)} - E(t)\cdot\sum_l \left[\boldsymbol{d}_{ml}^{k(t)}\rho_{ln}^{k(t)} - \rho_{ml}^{k(t)}\boldsymbol{d}_{ln}^{k(t)}\right] \qquad (2)$$

where $\varepsilon_n^k$ is the energy of band $n$ at momentum $k$, $T_2$ is the phenomenological dephasing time, $d$ is the transition dipole moment operator, $\rho^k$ is the density matrix at momentum $k$, $\delta_{mn}$ is the Kronecker delta. The SBEs were built upon the Wannier-interpolated band structures and transition dipole moments obtained from the DFT and Wannier90 calculations.

The linearly polarized driving laser was modeled with a cos² envelope ($f(t) = cos^2(\pi t/\tau_P)$ for $-\tau_P/2 \le t \le \tau_P/2$, and 0 otherwise, where $\tau_P$ is the total pulse duration), and a full-width at half-maximum (FWHM) duration of two optical cycles (approximately 32 fs for a 4800 nm laser). The peak intensity of the laser pulse corresponds to the material's damage threshold at each specific pressure.

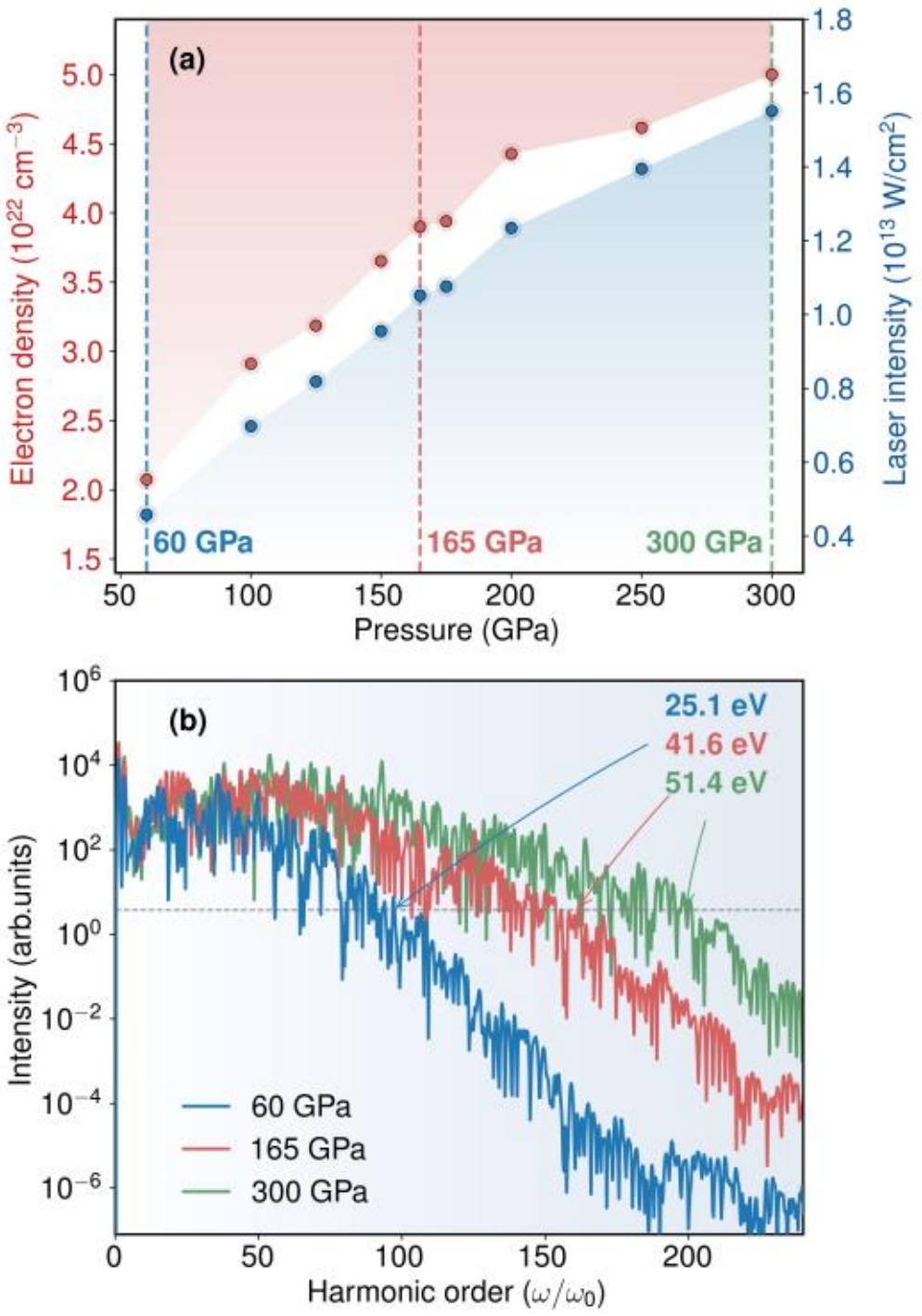

FIG. 2. Pressure dependent critical electron density and laser damage threshold of $h$-$WN_6$, and simulated high harmonic spectra at corresponding thresholds. (a) Critical electron density (left axis, red) and laser induced damage threshold intensity (right axis, blue) as a function of pressure in the range of 60–300 GPa. (b) Simulated harmonic spectra of $h$-$WN_6$ at 60 GPa, 165 GPa, and 300 GPa. The HHG is driven by the same laser parameters with the damage threshold intensities extracted from the top panel, namely $4.7\times10^{12}$ W/cm$^2$, $1.1\times10^{13}$ W/cm$^2$, and $1.6\times10^{13}$ W/cm$^2$.

By repeating this procedure for different pressure values, we calculate the critical electron density that $h$-$WN_6$ can sustain before entering into regime of lattice instability. As shown in Fig. 2(a), the critical electron density exhibits a continuous increase from $2.1\times10^{22}$ cm$^{-3}$ at 60 GPa to $5.0\times10^{22}$ cm$^{-3}$ at 300 GPa. This indicates that at higher pressures, the material can accommodate substantial photoexcited electrons before phonon modes soften. This increased tolerance to electronic excitation at higher pressure translates straightforwardly to a larger laser-induced damage threshold. Fig. 2(a) also shows the calculated damage threshold intensity as a function of pressure, for $h$-$WN_6$ irradiated by an intense linearly polarized, mid-infrared (4800 nm) and $\cos^2$-enveloped laser pulse with two optical cycles. The threshold intensity rises from $4.7\times10^{12}$ W/cm$^2$ at 60 GPa to a significantly larger value of $1.6\times10^{13}$ W/cm$^2$ at 300 GPa. This more than three-fold increase in the tolerable laser intensity is a crucial factor enabling the exploration of HHG in stronger driving fields.

The HHG spectra were simulated using linearly polarized light with the electric field oriented along the Γ-K direction of the Brillouin zone. The SBE simulations also provided the time evolution of electron populations in different bands. The total macroscopic current density $j(t)$ is the source of the harmonic emission and includes both intraband ($\boldsymbol{j}_{\text{intra}}$) and interband ($\boldsymbol{j}_{\text{inter}}$) contributions. The total intensity of high-harmonic spectrum can be obtained by $S(\omega) \propto \omega^2 |J(\omega)|^2$ , where $J(\omega)$ is the Fourier transform of the total current density, i.e., $J(\omega) = \mathcal{F}\left[\boldsymbol{j}_{\text{intra}}(t) + \frac{d}{dt}\boldsymbol{j}_{\text{inter}}(t)\right]$. The anisotropy of the HHG was investigated by rotating the laser polarization direction within the M-Γ-K plane during the SBE simulations.

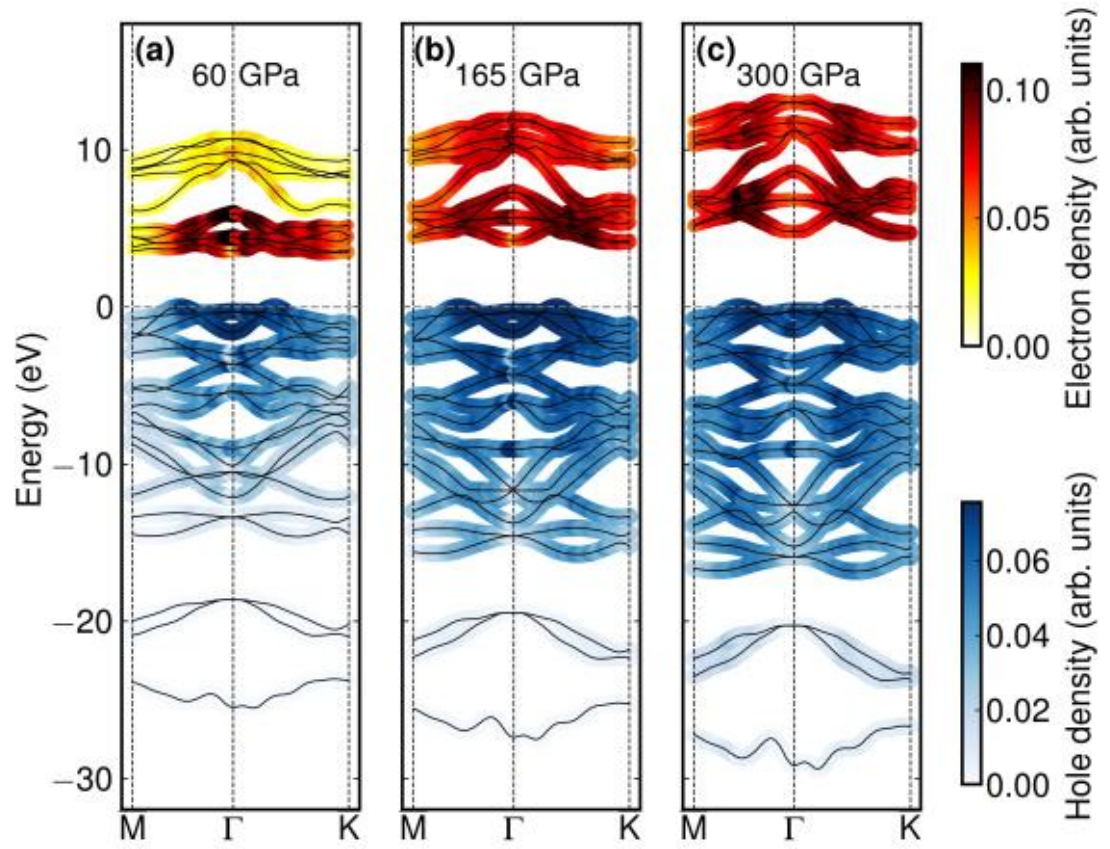

FIG. 3. Laser-induced electron and hole populations on energy bands of high-pressure $h$-$WN_6$. The calculated electron and hole populations are respectively projected onto conduction and valence bands along the M–Γ–K path, immediately after the laser pulse. (a), (b), and (c) correspond to pressures of 60 GPa, 165 GPa, and 300 GPa, respectively, with the system driven at its damage threshold intensity for each pressure.

The ability to apply higher laser intensities is a primary requirement for extending the HHG cutoff energy. We perform SBE simulations using the laser damage threshold intensities at different high pressures. The calculated HHG spectra for $h$-$WN_6$ at pressures of 60 GPa, 165 GPa, and 300 GPa are presented in Fig. 2(b). At 60 GPa, a clear harmonic plateau up to the 97th order is observed, with a cutoff energy around 25.1 eV. As the pressure increases to 165 GPa, the cutoff extends significantly to approximately 41.6 eV (161st order harmonic). Remarkably, at 300 GPa, the HHG cutoff reaches approximately 51.4 eV (199th order harmonic). This represents a more than two-fold increase in cutoff energy compared to that at 60 GPa.

To understand the mechanism behind this dramatic cutoff extension, we analyzed the laser-excited electron population in the conduction bands of $h$-$WN_6$ right after interaction with the intense driving laser. At 60 GPa shown in Fig. 3(a), the excited electron density is predominantly concentrated in the lower conduction bands. As the pressure increases to 165 GPa (Fig. 3(b)), there is a clear enhancement in the overall electron population within the conduction bands, and notably, a non-negligible fraction of electrons occupies higher-lying conduction bands. At 300 GPa, this trend becomes even more pronounced, with a substantial electron population found in all considered conduction bands across the whole Brillouin zone, as shown in Fig. 3(c). Thus, it demonstrates that the accessibility of high-energy electronic states can be improved at higher pressures under strong-field excitation, facilitating more energetic recombination pathways essential for extending the harmonic cutoff.

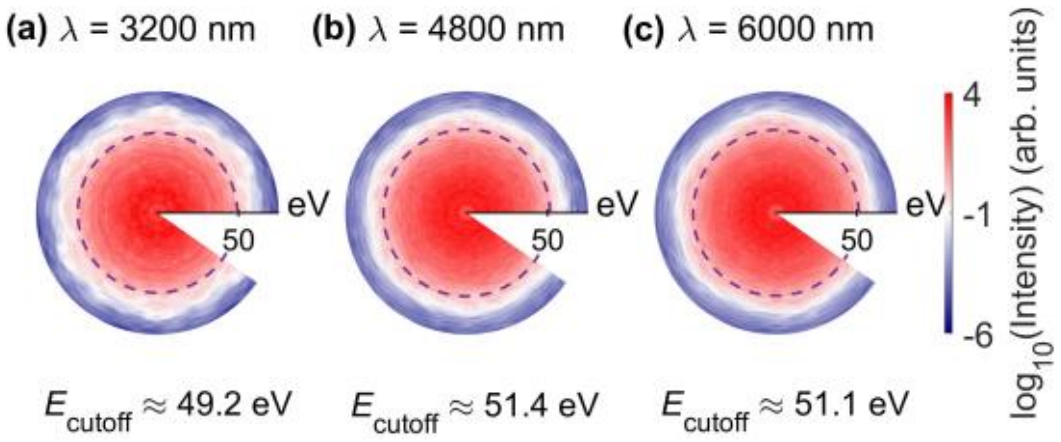

FIG. 4. Isotropic harmonic emission from $h$-$WN_6$ at 300 GPa. Polar plots show the orientation dependence of simulated HHG spectra driven by (a) 3200 nm, (b) 4800 nm, and (c) 6000 nm laser with corresponding damage thresholds. The driving laser is linearly polarized within the M–Γ–K plane.

A crucial aspect for practical application is the robustness of the HHG with respect to experimental conditions. We investigate the orientation dependence of harmonic emission at 300 GPa by rotating the laser polarization within the M–Γ–K plane for various driving laser wavelengths. As shown in Fig. 4, the harmonic emission from $h$-$WN_6$ is highly isotropic for driving laser wavelengths of 3200 nm, 4800 nm, and 6000 nm, using corresponding damage threshold intensities. In all the cases, neither the emission efficiency nor the cutoff energy exhibits obvious orientation dependence. This pronounced isotropy, maintained across a broad range of mid-infrared wavelengths, substantially relaxes experimental constraints on crystal orientation and laser parameters, which is particularly beneficial for high-pressure HHG measurements where precise alignment is often challenging.

In conclusion, we theoretically predicted that the HHG in $h$-$WN_6$ can be significantly enhanced through a pressure-tuning strategy, which opens new possibilities for compact EUV sources achieving higher efficiencies and higher photon energies. It is noteworthy that, in this work, we employ an accurate Wannier model constructed from a limited subspace (see Supplemental Fig. S4). Higher-lying conduction bands are not included, due to intrinsic limitations of the Wannier-based self-consistent wave-function basis in reproducing extremely high-energy conduction states. As a result, our predictions are expected to be conservative. Moreover, comparative studies on cubic diamond and $h$-$MoN_6$ (see Supplemental Fig. S5) confirm similar pressure-induced enhancement of HHG in these superhard semiconductors. Access to coherent EUV radiation from compact sources is critical for advancing attosecond science (e.g., generating shorter attosecond pulses), enabling new frontiers in time-resolved spectroscopy, and facilitating higher-resolution imaging techniques. Our findings pave the way for exploring these applications using pressure-tuned solid-state platforms.

*jiansun@nju.edu.cn
†chaoyu@njust.edu.cn
‡rflu@njust.edu.cn
§These authors contributed equally to this work.

## Supporting Information

## Remarkable Enhancement of High Harmonic Generation from Superhard Material under High Pressure

Zishao Wang,[1,§] Tong Wu,[1,§] Ziwen Wang,[1,§] Sicheng Liu,[1] Hui Li,[1] Kun Zhao,[1] Jian Sun,[2,*] Chao Yu,[1,†] and Ruifeng Lu[1,‡]

To assess the finite-temperature structural stability of *h*-$WN_6$, we performed NPT ab initio molecular dynamics (AIMD) simulations for pristine *h*-$WN_6$ at 60, 165, and 300 GPa. For each pressure, simulations were carried out at both 300 and 500 K using a 168-atom supercell. The time step was 1 fs, and the total simulation time was approximately 10 ps.

The structural evolution was monitored using the relative total energy per atom, the mean-square displacement (MSD) of N atoms, and the nearest-neighbor N-N distances. Figure S1 shows that the change in total energy per atom, defined as $\Delta E = [E(t) - E_0]/n$ (where $E_0$ is the total energy of equilibrium structure and $n$ is the number of atoms in the constructed cell), fluctuates within a bounded range without abrupt jumps throughout the simulations. As shown in Figure S2, the MSDs remain bounded throughout the simulations at all investigated pressures and temperatures, indicating the absence of clear atomic diffusion within the simulated timescale. Figure S3 shows that the nearest-neighbor N-N distances fluctuate around stable values without abrupt elongation, indicating that the armchair-like N6 ring units remain intact. No structural collapse or transformation is observed during the simulations. These AIMD results support the finite-temperature structural integrity of *h*-$WN_6$ under the pressure and temperature conditions considered in this work.

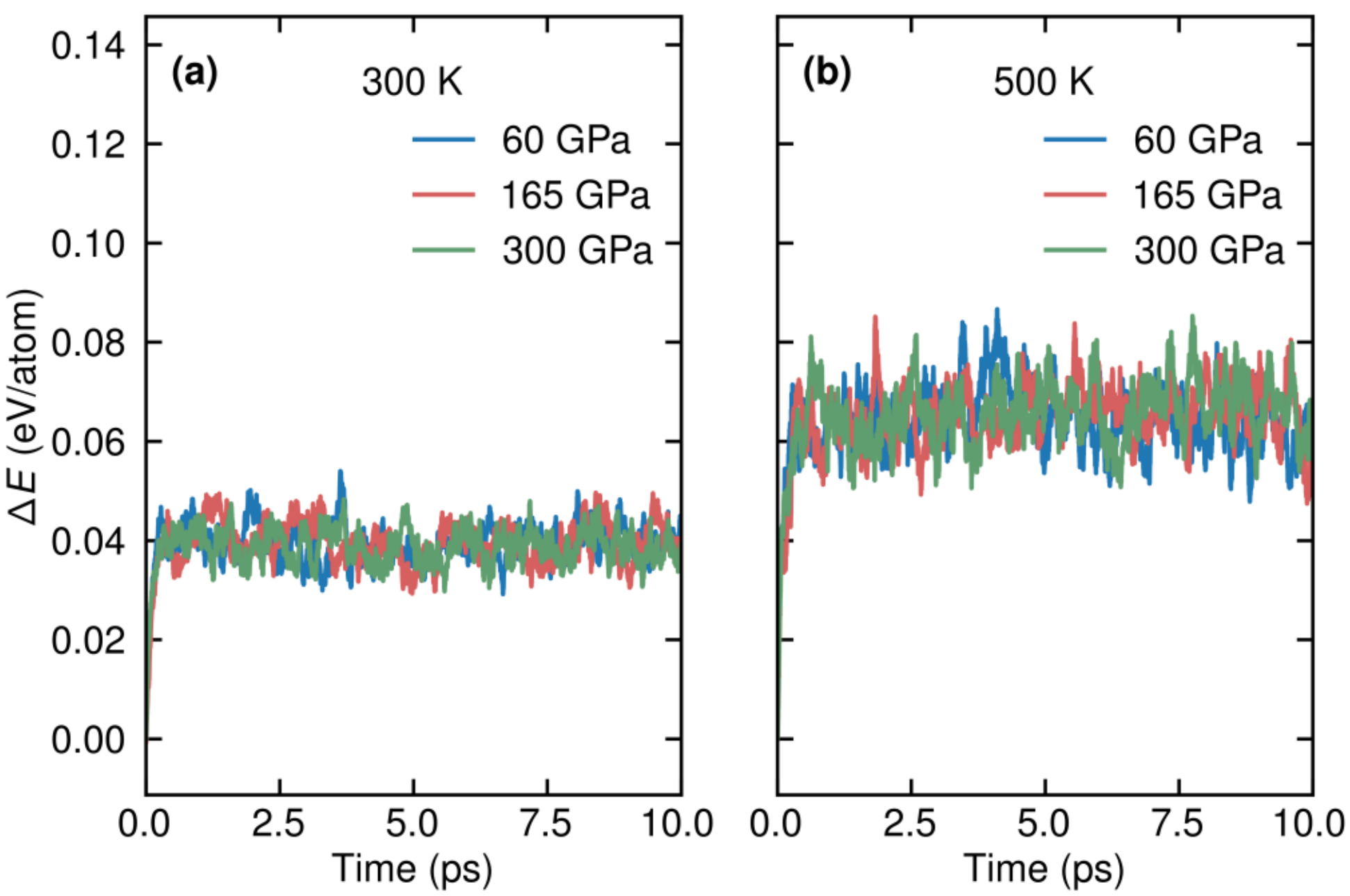

**Figure S1.** Time evolution of the relative total energy per atom in $h$-$WN_6$ during NPT AIMD simulations.

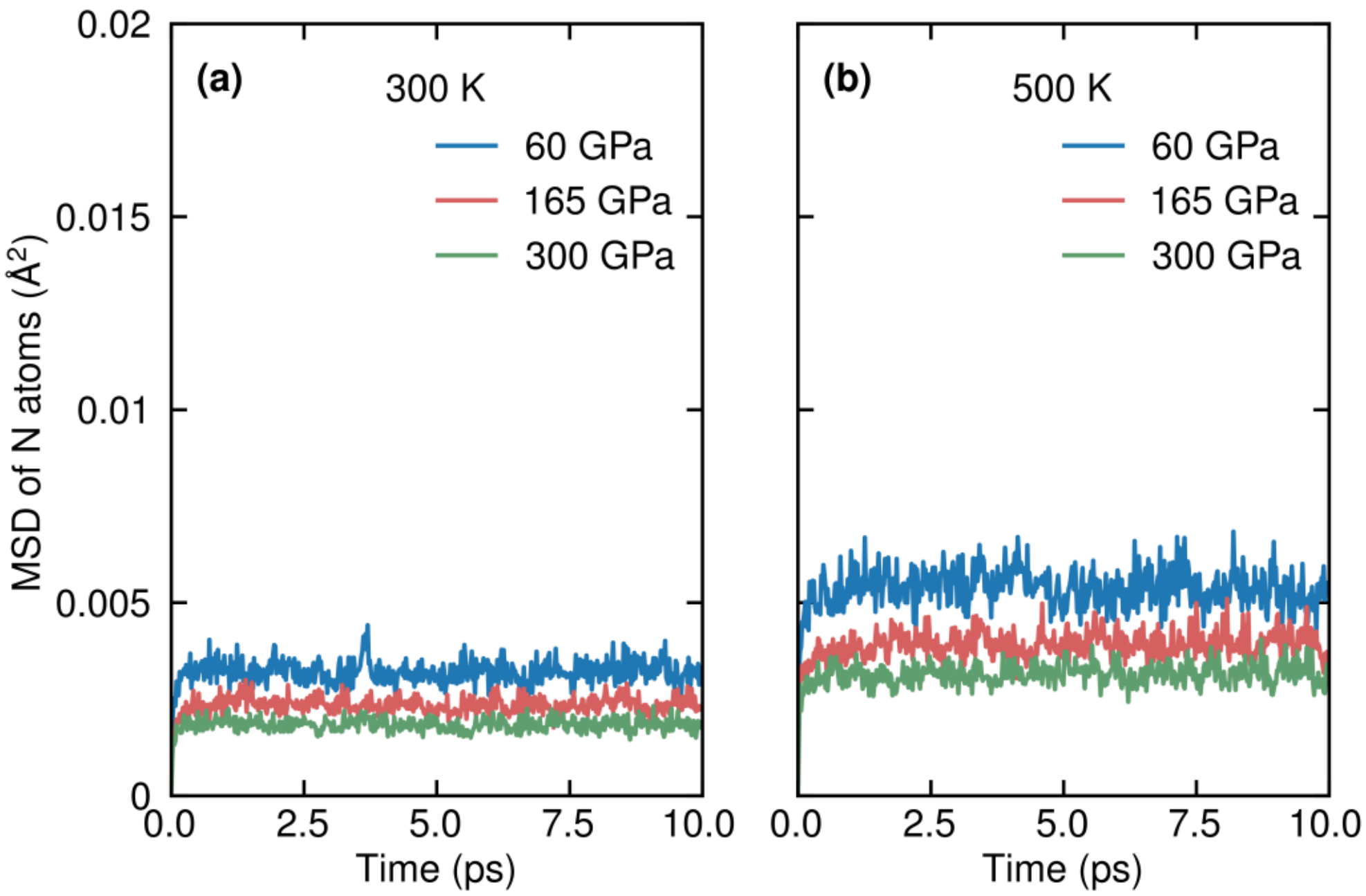

**Figure S2.** Mean-square displacements of N atoms in $h$-$WN_6$ during NPT AIMD simulations.

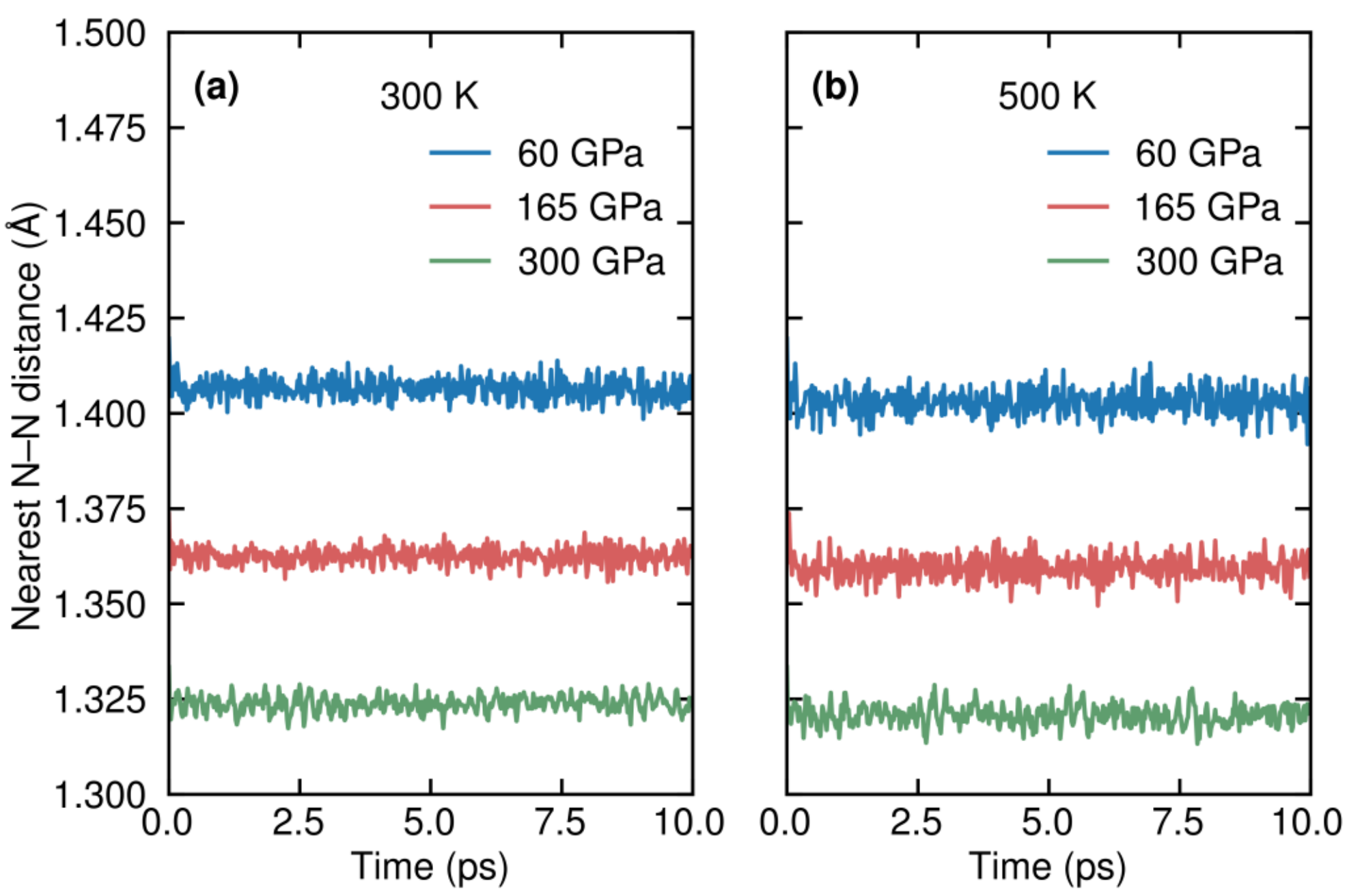

**Figure S3.** Time evolution of the nearest neighbor N-N distance in *h*-$WN_6$ during NPT AIMD simulations.

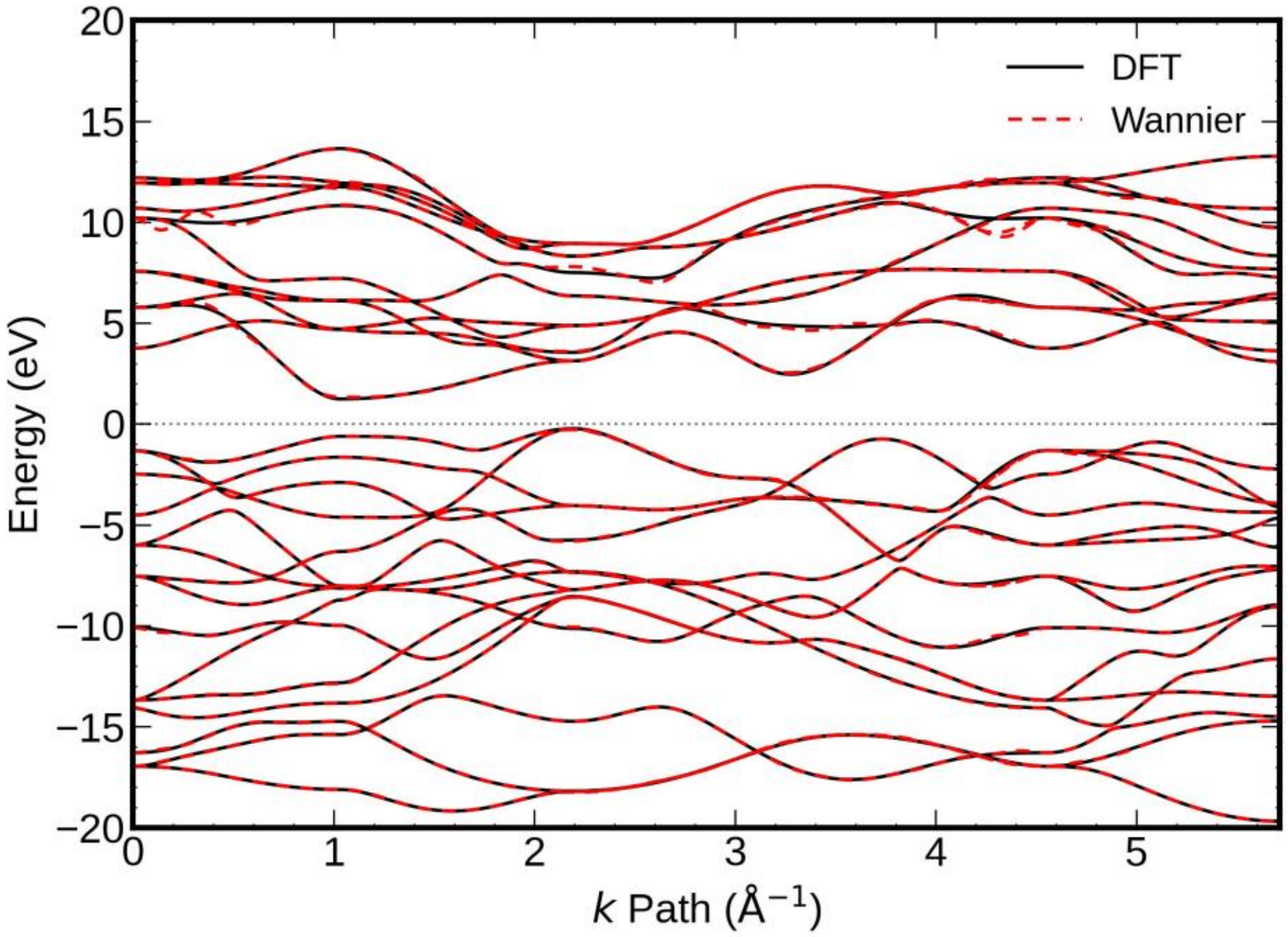

**Figure S4.** Comparison between the DFT-calculated electronic band structure (black solid lines) and the Wannier-interpolated bands (red dashed lines) for *h*-$WN_6$ at 300 GPa. The excellent agreement within the 29-band window confirms the accuracy of the Hamiltonian used in the SBE simulations. It should be noted that the Wannier model used to construct the SBE basis has intrinsic limitations in reproducing very high-lying conduction bands.

For cubic diamond, increasing pressure from 0 to 300 GPa (band gap widening from 4.02 to 4.88 eV) raises the critical electronic excitation percentage from 5.8% to 6.7%. Using an 800 nm laser with a FWHM duration of 10 optical cycles (~27 fs), the damage threshold changes from $9.6\times10^{12}$ W/cm$^2$ at 0 GPa to $2.0\times10^{13}$ W/cm$^2$ at 300 GPa, and the cutoff extends clearly from the 11th to the 17th harmonic order, as shown in Figure S5(a).[1] Similarly, for *h*-$MoN_6$ (a structural analog of *h*-$WN_6$), increasing pressure from 0 to 50 GPa (band gap widening from 0.37 to 0.58 eV) raises the critical electronic excitation percentage from 1.2% to 4.7%. Using a 4800 nm laser with a FWHM duration of 2 optical cycles (~32 fs), the damage threshold changes from $6.0\times10^{11}$ W/cm$^2$ at 0 GPa to $3.6\times10^{12}$ W/cm$^2$ at 50 GPa, and the cutoff extends clearly from the 31st to the 67th harmonic order, as shown in Figure S5(b).[2]

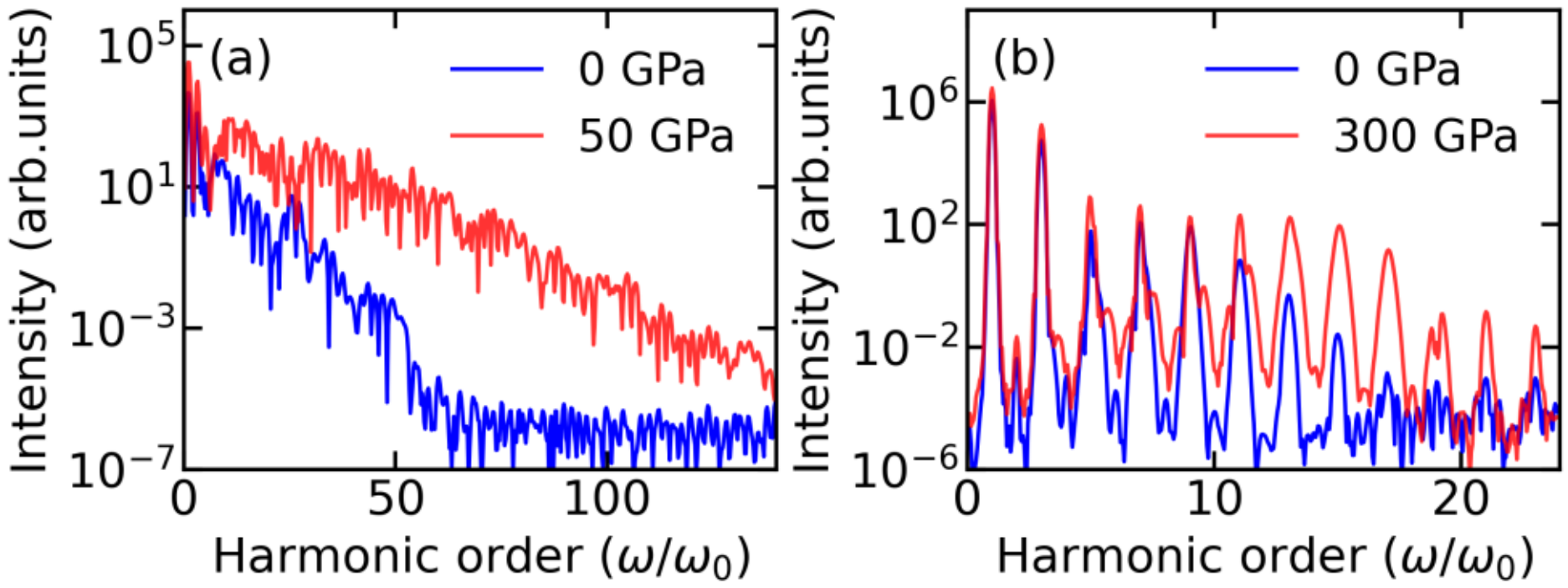

**Figure S5.** HHG spectra of (a) cubic diamond at 0 and 300 GPa and (b) *h*-$MoN_6$ at 0 and 50 GPa. The spectra were driven by laser pulses with peak intensities corresponding to the damage thresholds at each pressure.

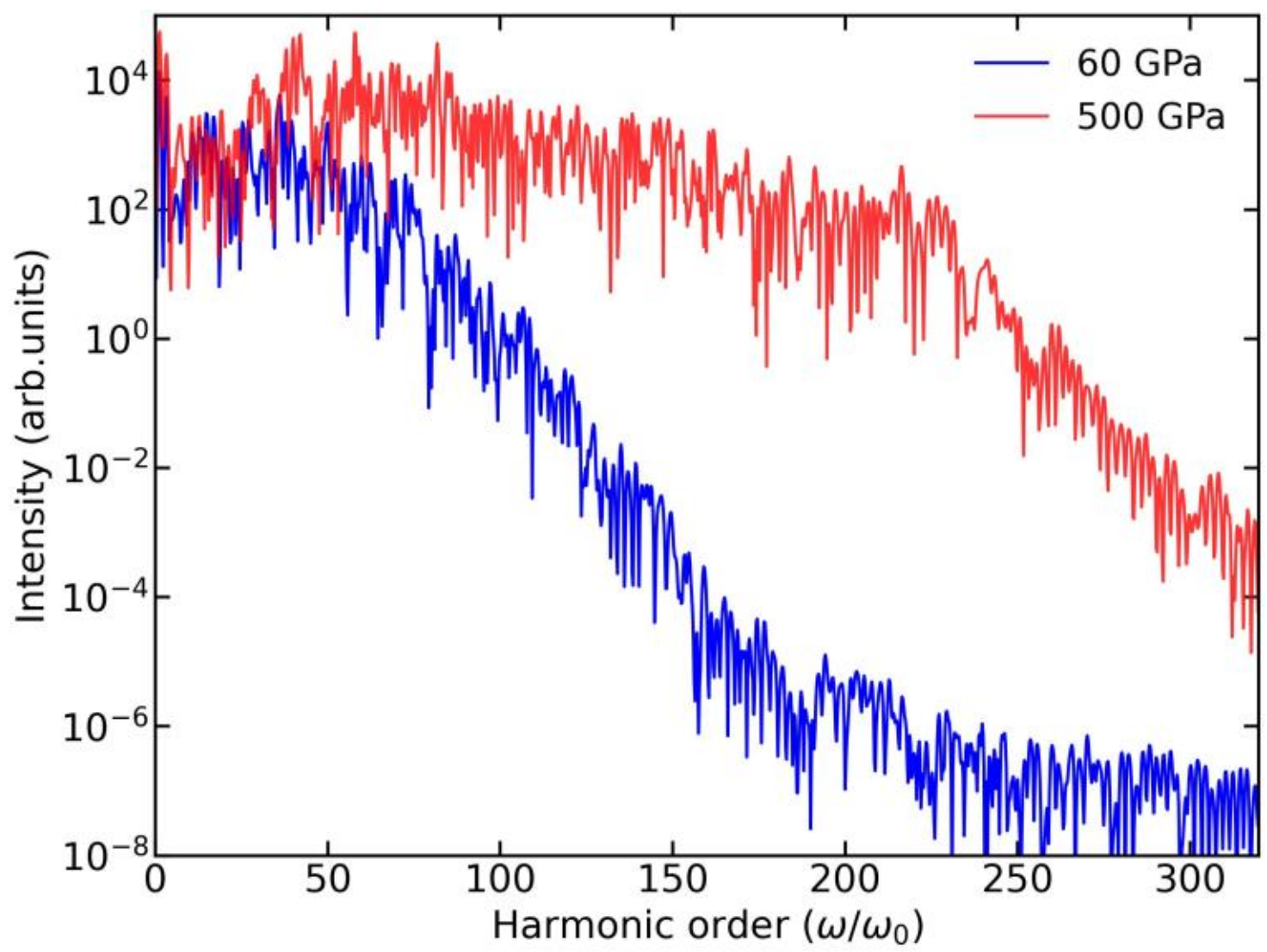

**Figure S6.** High harmonic generation spectra of *h*-$WN_6$ at 60 GPa and 500 GPa. The HHG is driven by the same laser parameters with the damage threshold intensities $4.7\times10^{12}$ W/cm$^2$ at 60 GPa and $2.4\times10^{13}$ W/cm$^2$ at 500 GPa.